\documentclass{article}
\usepackage{pgfplots}
\usepackage{lipsum}
\usepackage{caption}
\usepackage{subcaption}
\usepackage{authblk}
\usepackage[english]{babel}
\usepackage{amsfonts}
\usepackage{amsmath}
\usepackage[top=2cm, bottom=2cm, left=2cm, right=2cm]{geometry}
\usepackage{fancyhdr}
\usepackage{multicol}
\usepackage{bbold}
\usepackage{epsfig}
\usepackage{graphics}
\usepackage{epsfig}
\usepackage{lipsum}
\usepackage{fancyhdr}
\usepackage{booktabs}
\usepackage{wrapfig}
\usepackage[export]{adjustbox}
\usepackage{rotating}
\usepackage{amsmath}

\usepackage{hyperref}
\usepackage{cleveref}

\usepackage{cite}
\renewenvironment{abstract}{\hfill\begin{minipage}{0.95\textwidth}
		\rule{\textwidth}{1pt}}
	{\par\noindent\rule{\textwidth}{1pt}\end{minipage}}

\makeatletter
\usepackage{orcidlink}
\makeatother

\begin{document}
\title{Constructing Multipartite Planar Maximally Entangled States from Phase States and Quantum Secret Sharing Protocol}

\author[1]{\textbf{L. Bouhouch}}
\author[1]{\textbf{Y. Dakir \normalsize\orcidlink{0009-0005-3408-1309}}}
\author[1,2]{\textbf{A. Slaoui \normalsize\orcidlink{0000-0002-5284-3240}}{\footnote {Corresponding author: {\sf abdallah.slaoui@fsr.um5.ac.ma}}}}
\author[1,2]{\textbf{ R. Ahl Laamara \normalsize\orcidlink{0000-0002-8254-9085}}}
	\affil[1]{\small LPHE-Modeling and Simulation, Faculty of Sciences, Mohammed V University in Rabat, Rabat, Morocco.}
	\affil[2]{\small  Centre of Physics and Mathematics, CPM, Faculty of Sciences, Mohammed V University in Rabat, Rabat, Morocco.}
	\maketitle
\begin{abstract}
In this paper, we explore the construction of Planar Maximally Entangled (PME) states from phase states. PME states form a class of $n$-partite states in which any subset of adjacent particles whose size is less than or equal to half the total number of particles is in a fully entangled state. This property is essential to ensuring the robustness and stability of PME states in various quantum information applications. We introduce phase states for a set of so-called noninteracting $n$ particles and describe their corresponding separable density matrices. These phase states, although individually separable, serve as a starting point for the generation of entangled states when subjected to unitary dynamics. Using this method, we suggest a way to make complex multi-qubit states by watching how unconnected phase states change over time with a certain unitary interaction operator. In addition, we show how to derive PME states from these intricate phase states for two-, three-, four-, and K-qubit systems. This construction method for PME states represents a significant advance over absolutely maximally entangled (AME) states, as it provides a more accessible and versatile resource for quantum information processing. Not only does it enable the creation of a broader class of multipartite entangled states, overcoming the limitations of AME states, notably their restricted availability in low-dimensional systems; for example, the absence of a four-qubit AME state, but it also offers a systematic construction method for any even number of qudits, paving the way for practical applications in key quantum technologies such as teleportation, secret sharing and error correction, where multipartite entanglement plays a central role in protocol efficiency.

	\end{abstract}
	
	\vspace{0.5cm}
	%\textbf{Keywords}: Multiparameter estimation, Photon-Added coherent states, Individual and simultaneous strategies.
%This method of constructing PME states is particularly relevant for applications in fields such as quantum teleportation, quantum secret sharing, and quantum error correction, where multiparty entanglement plays a central role in the efficiency of the protocols.

\section{Introduction}
Entanglement has been a prominent topic since the inception of quantum mechanics, sparking significant discussions, most notably the Einstein-Podolsky-Rosen (EPR) paradox \cite{Einstein1935,Nielsen2010}, which ultimately inspired Bell to develop a method for measuring entanglement \cite{Bell1964, Bartkiewicz2013, Slaoui2023,Amghar2023,Horodecki2009,Guhne2009}. While the classification of entanglement classes for three- and four-qubit systems is now well-established, and canonical forms of pure states under local unitary transformations in each local Hilbert space have been extensively analyzed \cite{Brierley2007,Ghahi2016}, our understanding of entanglement diminishes as the number of local quantum degrees of freedom increases. This is due to the exponential growth of the number of independent invariants classifying entanglement, making the purpose of each category increasingly ambiguous. However, significant progress has been made in classifying maximally multipartite entangled states composed of qubits \cite{Osterloh2006,Gour2010}. New classifications of multipartite entanglement are presented in Refs \cite{Gharahi2020,Gharahi2021}, the former resembling a Mendeleev table and using “k-sequential families”, while the latter uses algebraic geometry to classify three-qudits entanglement. Ref.\cite{Gharahi2020} quantifies entanglement, potentially serving as a monotone, and provides information on the complexity of the state, the identification of the relevant state and the construction of the entanglement witness. Ref.\cite{Gharahi2021} proposes a detailed three-qubit classification, forming a kernel for higher-dimensional systems, and uses tensor rank to quantify the strength of entanglement, with implications for SLOCC conversions and mixed-state analysis.\par  

The characterization of multipartite systems has important applications in quantum information, including quantum teleportation \cite{Bennett1993,Bouwmeester1997,Dakir2023,Slaoui2024}, quantum cryptography \cite{Bennet1984, Ekert1991}, and quantum computation \cite{nielsen2000, Barenco1995}. However, it is still hard to measure quantum correlations in multipartite quantum systems. Different methods have been used in the past to look at important aspects of entangled multiqubit states \cite{Gerke2015,Li2015,Zhang2016,Assadi2016,Pezze2017}. Given that entanglement is a fundamental resource for many applications in quantum information, quantifying the degree of entanglement in any multi-qudit system is essential. Significant strides have been made in categorizing maximally multipartite entangled states of qubits \cite{Coffman2000,Ganczarek2012,Helwig2014}. In a composite multipartite system, there exist several different classes of multipartite entangled states \cite{Carvalho2004,Love2007,Meyer2002,Qun2014,Scott2004}. One important way to describe entanglement in such systems is by considering some of its bipartitions and then averaging a certain entanglement measure. Multi-qubit entangled states arise when qubits in a multi-qubit system interact. This interaction is modeled by applying the Ising-type unitary operator $\mathcal{C}_{ij}$ acting on qubits $i$ and $j$. Many studies have been done on different important aspects of entangled multi-qubit states \cite{Gour2010}. Also, many studies have been done on multipartite entanglement \cite{Haddadi2018,Akhound2019,Pezze2017}.\par

In recent years, researchers have studied the concept of absolutely maximally entangled (AME) states to gain a complete understanding of multipartite entangled states. These states are significant in various areas of quantum information, such as teleportation and quantum secret sharing, and provide connections between different fields of mathematics. AME multipartite states, are maximally entangled for any possible bipartition of the multi-qudit system, they are states of $k$ particles such that any collection on $\lfloor \frac{k}{2}\rfloor$ particles are in maximally mixed states. For each fixed number of particles, AME states are powerful in high-dimensional states, but severely restricted in low dimensions. For instance, AME states do not exist for 7 qubits or 4 qubits. This necessitates the search for a new, wider class of entangled states with fewer constraints or, ideally, no constraints at all. Generalizations of AME states, known as PME states, have been studied \cite{karimipour2020}.\par

On the other hand, there are a variety of reasons to communicate with a valid receiver while ensuring their privacy. Current communication protocols are vulnerable to technological advancements and computationally demanding attacks due to their reliance on the difficulty of breaking them. On the other hand, quantum protocols provide unwavering security based on the laws of physics by utilizing the fundamental ideas of quantum physics, independent of the attacker's computing power. Quantum Secret Sharing (QSS), initially introduced by Hillery using a three-particle GHZ state, is one of the most secure quantum protocols \cite{Hillery1999}. Once the threshold QSS system reaches a predetermined number of participants, only a subset of players can access the secret. The increasing significance of quantum security in real-world applications is evidenced by the numerous additional quantum secret sharing techniques that have been put forth in the literature \cite{Tittel1999}. Furthermore, absolutely maximum entangled states are used in QSS protocols, illustrating how versatile these states are for applications involving quantum secret sharing \cite{Helwig2013}. Multiple significant studies have experimentally validated QSS, making it a tried-and-true technology for security applications.\par

The PME states are a versatile construct that can be formulated for any dimensional space and for any specified number of particles, collectively referred to as PME \((K,d)\) states \cite{karimipour2020}. A defining characteristic of these states is that any collection of \(\lfloor \frac{k}{2}\rfloor\) adjacent particles, which are subjected to periodic boundary conditions, exists in a completely mixed state. This property plays a crucial role in their applicability across various quantum information protocols. Specifically, PME states are particularly well-suited for implementations in parallel teleportation \cite{Bennet1984}, quantum secret sharing \cite{Shamir1979}, and quantum error correction \cite{Shor1995}. PME states are flexible because they can keep their strong structure even when faced with different restrictions. This makes quantum computing and communication possible. Because they are completely mixed with particles next to each other, they can effectively send and encode information while being as resistant to noise and interference as possible. This feature makes them a powerful tool in enhancing the efficiency of quantum networks. Moreover, the application of PME states in quantum secret sharing highlights their capability to secure and distribute information among multiple parties, ensuring that only designated recipients can access the shared secrets. In quantum error correction, their properties make it possible to create protocols that can find and fix mistakes that happen during quantum computations or communications, keeping the integrity of the quantum information safe. Overall, the exploration of PME states opens new avenues for research and application in quantum technologies, offering promising strategies for advancing the field of quantum information science.\par

In this work, the central idea concerns the construction of PME states from entangled phase states of multipartite qudit system. In section \ref{sec2} we started by presenting the  basics about PME states and giving their various and necessary definitions.  In section \ref{sec3} we discuss the $k$-qubits algebra generated by operators  $k$ qubits denoted $ \{ \mathcal{B}_{i}^+, \mathcal{B}_{i}^- , \mathcal{M}_i; \quad i=1,2,\cdots,k\}$ this algebra is a particular form of Weyl Heisenberg algebra, by using the polar decomposition of the raising and lowering operators associated with $k$ qubits, the phase operators was defined and solving corresponding eigenvalue equation enable us to derive phase states. Using the factorized form of the evolution operator the evolved phase states was obtained in section \ref{sec4}. The PME states from entangled  phase states for $k=2,3,4,8, K$ qudit systems are investigated in section \ref{sec5}. Then we showed in section\ref{sec6}, how a PME state can be used to construct a quantum secret sharing (QSS) that allows distant parts to encode the secret and share it from one part to another that make quantum teleportation one of the strangest uses of entanglement, Finally we close this paper with concluding remarks in section \ref{clc}.

\section{Basics about PME states} \label{sec2}
Planar maximally entangled states were introduced to address certain limitations associated with AME states \cite{karimipour2020}. These states are characterized by the fact that each half of adjacent particles, forming connected subsets, exists in a fully mixed state. This feature suggests the possibility of a significantly larger number of PME states compared to AME states.
\begin{itemize}
	\item PME states are $k$-partite states in which each subset of adjacent particles, of size less than or equal to $\lfloor \frac{k}{2}\rfloor$ is in a totally mixed state.
	\item For $k \le 3$, PME states are identical to the AME states. Taking into account the periodic boundary conditions, there is no difference between them,
	\begin{equation}
	    \left| {\mathcal{PME}} \right\rangle  = \left| {\mathcal{AME}} \right\rangle  = \frac{1}{{\sqrt {{d^n}} }}\sum\limits_{\mu \in \mathbb{Z}_d^n} {{{\left| {{\mu_1}} \right\rangle }_{{\mathcal{B}_{1}}}}...{{\left| {{\mu_n}} \right\rangle }_{{\mathcal{B}_{n}}}}{{\left| {\phi (\mu)} \right\rangle }_a}}.
	\end{equation}
	\item For PME states with $k \ge 4$ to AME states. The application involves a self-controlled phase gate operation. This operation keeps the entanglement of two partitions for a qudit within connections, focusing on structural compatibility while adding phase changes that don't affect the level of entanglement.
 
\end{itemize}
To construct various families of PME states in different dimensions, we consider this initiatory that a partition of the particles into parts ${\mathcal{A}_{1}}$ and ${\mathcal{A}_{2}}$ with $\left| {{\mathcal{A}_{1}}} \right| \le \left| {{\mathcal{A}_{2}}} \right|$ and the corresponding state $\left| \xi  \right\rangle $  given by
\begin{equation}  \mathord{\buildrel{\lower3pt\hbox{$\scriptscriptstyle\frown$}}\over \xi } : = \sum\limits_\mu {{{\left| {{\phi _\mu}} \right\rangle }_{{\mathcal{A}_{1}}}}} {\left\langle \mu \right|_{{\mathcal{A}_{2}}}}{\kern 1pt} {\kern 1pt} {\kern 1pt} {\kern 1pt} {\kern 1pt} :{\kern 1pt} {\kern 1pt} {\kern 1pt} {\kern 1pt} {\kern 1pt} {H_{{\mathcal{A}_{1}}}} \to {\kern 1pt} {\kern 1pt} {\kern 1pt} {\kern 1pt} {\kern 1pt} {\kern 1pt} {H_{{\mathcal{A}_{2}}}}
\end{equation}
where, $\mathord{\buildrel{\lower3pt\hbox{$\scriptscriptstyle\frown$}}\over \xi } $ is an PME state if the density matrix ${\varrho _{{\mathcal{A}_{1}}}}$ of part ${{}}$ becomes maximally mixed i,e $\mathord{\buildrel{\lower3pt\hbox{$\scriptscriptstyle\frown$}}\over \xi } $ be proportional to isometry, that is ${{\mathord{\buildrel{\lower3pt\hbox{$\scriptscriptstyle\frown$}}\over \xi } }^ + }\mathord{\buildrel{\lower3pt\hbox{$\scriptscriptstyle\frown$}}\over \xi }  \propto \mathcal{I}$ this condition should be satisfied only for connected partitions under the periodic boundary condition. 
When $\left| {{\mathcal{A}_{1}}} \right| = \left| {{\mathcal{A}_{2}}} \right| = K$, the operator $\mathord{\buildrel{\lower3pt\hbox{$\scriptscriptstyle\frown$}}\over \xi }$ is proportional to the identity and the density operators of both parts ${\mathcal{A}_{1}}$ and  ${\mathcal{A}_{2}}$ given as ${\varrho _{{\mathcal{A}_{1}}}} = \frac{1}{K}{\mathcal{I}_{{\mathcal{A}_{1}}}}$ and also ${\varrho _{{\mathcal{A}_{2}}}} = \frac{1}{K}{\mathcal{I}_{{\mathcal{A}_{2}}}}$.\\

Furthermore it is shown in \cite{karimipour2020} that the PME state $\left| \xi  \right\rangle$, shared between any division of the
$\left| {{\mathcal{A}_{1}} + {\mathcal{A}_{2}}} \right|$  qudits into connected subsets of equal size $|{\mathcal{A}_{1}}| = |{\mathcal{A}_{2}}|$ qudits, is giving by
\begin{equation}
	\left| \xi  \right\rangle  = \frac{1}{{\sqrt {{d^{\left| {{\mathcal{A}_{1}}} \right|}}} }}\sum\limits_{{\mathcal{A}_{1}},{\mathcal{A}_{2}}} {{\xi _{{\mathcal{A}_{1}},{\mathcal{A}_{2}}}}} \left| {{\mathcal{A}_{1}}} \right\rangle \left| {{\mathcal{A}_{2}}} \right\rangle,
\end{equation}
where ${{\xi _{{\mathcal{A}_{1}},{\mathcal{A}_{2}}}}}$ is the matrix representation of the state coefficients, then a PME state of $2k$ qudits
\begin{equation}
	\left| \xi  \right\rangle  = \frac{1}{{\sqrt {{d^{\rm{k}}}} }}{\xi _{{i_1},{i_{2k}}}}\left| {{i_1},{i_2},....,{i_{2{\rm{k}}}}} \right\rangle.
\end{equation}
\textbf{Definition I:} We say that $\left| \xi  \right\rangle $ is PME state of $K$-qudits if every bipartition of $ \left\{ {1,......,K} \right\} $ of the systems into the sets ${\mathcal{A}_{2}}$ and ${\mathcal{A}_{1}}$ with $ n = \left| {{\mathcal{A}_{1}}} \right| \le \left| {{\mathcal{A}_{2}}} \right| = K - n $, , satisfies the following conditions:

1. The density matrix derived from $\left| \xi  \right\rangle \left\langle \xi  \right|$ tracing out the  adjacent and connected sites on the entries in ${\mathcal{A}_{1}}$ is proportional to the identity:
\begin{equation}
    {\varrho _{{\mathcal{A}_{2}}}} = T{r_{{\mathcal{A}_{1}}}}\left( {\left| \xi  \right\rangle \left\langle \xi  \right|} \right) = \frac{1}{{{d^n}}}{\mathcal{I}_{{d^n}}}.
\end{equation}

2. The entropy of every bipartition of $n$ local degree of freedom is maximal: 
\begin{equation}
    \left( {{\varrho _{{\mathcal{A}_{2}}}}} \right) = n{\log _2}d.
\end{equation}
\textbf{Definition II:} For $K$ even dimension $(K = 2k)$ of particles, the PME states shared between $2k$ parties  is given by
\begin{equation}
	\left| {\mathcal{PME}\left( {K,d} \right)} \right\rangle  = {\prod\limits_{l = 1}^k {\left| {{\mathcal{P}^ + }} \right\rangle } _{l,l + k}} = \prod\limits_{l = 1}^k {{{\left( {\frac{1}{{\sqrt d }}\sum\limits_{i = 0}^{d - 1} {\left| {i,i} \right\rangle } } \right)}_{l,l + k}}}.
\end{equation} \label{def2}

\section{Phase states of multi-qudit systems} \label{sec3}
A mathematical framework for describing quantum systems with conjugate variables, such as position and momentum, is provided by the Weyl-Heisenberg algebra \cite{Wong2000,Torresani1991,Dehdashti2013,Dehdashti2014}. Within this framework, phase operators play an essential role in defining phase states, which are quantum states with well-defined phase properties. They can be physically illustrated by a simple example involving a harmonic quantum oscillator. The Weyl-Heisenberg algebra for this system is generated by the position operator $X$ and the momentum operator $P$, which satisfy the canonical commutation relation $[X, P]=i\hbar$. From these operators, we can define the phase operator $\mathcal{B}$ as $\mathcal{B} = \arctan(X/P)$. The phase states $\left| \psi\right\rangle$ are then defined as eigenstates of this operator: $\mathcal{B}\left| \psi\right\rangle=\phi \left| \psi\right\rangle$, where $\phi$ represents the eigenvalue associated with the phase state. Physically speaking, these phase states correspond to quantum states where the phase relationship between the position and impulse components is well defined. For example, for this harmonic oscillator, a phase state with $\phi=0$ means that position and pulse are in phase, while a phase state with $\phi=\pi/2$ indicates that pulse is $\pi/2$ ahead of position. Devices such as ion traps and photonic arrays are physical examples of phase states, illustrating their relevance to quantum operations.\par

In the context of the following, qudits are generalizations of qubits and can be implemented using quantum physical systems $d$ levels $(d > 2)$. Consider the qudit algebra \( \mathcal{H} \) spanned by the operators \( \{ \mathcal{B}_{+}, \mathcal{B}_{-}, \mathcal{M} \} \) and the identity \( \mathcal{I} \), which satisfy the structural relations,
\begin{equation}
	[ \mathcal{B}_{-} , \mathcal{B}_{+} ] = (d-1)\mathcal{I} - 2\mathcal{M} , \qquad   [\mathcal{B}_{+} , \mathcal{M}] =  \mathcal{B}_{+}, \qquad   [\mathcal{B}_{-} , \mathcal{M} ] =  - \mathcal{B}_{-}. \label{quditalgebra}
\end{equation}
The qudit algebra \( \mathcal{H} \) is a variant of the generalized nonlinear Weyl-Heisenberg algebras \( \mathcal{A}_{k} \). Consequently, the corresponding representation space of the qudit algebra \( \mathcal{H} \) is \( d \)-dimensional, represented by the set \( \{ | \mu \rangle, ~ \mu = 0, 1, \ldots, d-1 \} \). This set is spanned by the eigenstates of the number operator $\mathcal{M}$, satisfying the relation \( \mathcal{M} | \mu \rangle = \mu | \mu \rangle \). The actions of the lowering operator $\mathcal{B}_{-}$ and the raising operator $\mathcal{B}_{+}$ on the basis of the Hilbert space \( \mathcal{H} \) are defined as follows 
\begin{equation}
	\mathcal{B}_{+} \vert \mu \rangle = \left(\mathcal{F}(\mu+1)\right)^{1/2}~ \vert \mu+1 \rangle, \qquad \mathcal{B}_{-} \vert \mu \rangle = \left(\mathcal{F}(\mu)\right)^{1/2}~ \vert \mu-1 \rangle,
\end{equation}
\begin{equation}
	\mathcal{B}_{-} \vert  0 \rangle = 0  \qquad \mathcal{B}_{+} \vert  d-1 \rangle = 0.
\end{equation}
with $\mathcal{F}(k)$ is the structure function defined by  $\mathcal{F}(\mu) = \mu(d-\mu)$. The creation and annihilation operators satisfy the following nilpotency conditions
\begin{equation}
    (\mathcal{B}_{-})^d = 0 \qquad (\mathcal{B}_{+})^d = 0,
\end{equation}
We now provide the number operators, raising, and lowering associated with $K$ qubits, which are represented by the notation $ \{ \mathcal{B}_{i}^+, \mathcal{B}_{i}^-, \mathcal{M}_i; \quad i=1,2,\cdots,K\}$. The commutation relations are satisfied by 
\begin{equation}\label{eq010}
	\left[ \mathcal{B}_{i}^-, \mathcal{B}_{j}^+\right ] = \bigg((d-1)\mathcal{I} - 2 \mathcal{M}_{i}\bigg)~ \delta_{ij}, \quad   \left[ \mathcal{B}_{i}^+, \mathcal{B}_{j}^+\right ] =  \left[ \mathcal{B}_{i}^-,
	\mathcal{B}_{j}^-\right ]=0,\quad \left[ \mathcal{B}_{i}^{\pm}, \mathcal{M}_j \right ]  = \pm \delta_{ij} \mathcal{B}_{i}^{\pm} .
\end{equation}
The multi-qubit Hilbert space associated with the $k$ non-interacting qubit systems is given as the tensor product of $k$ copies of the single-qubit Hilbert space ${\cal H}^{\otimes k}$ with the basis $\{ \vert \mu_1, \mu_2,\cdots,\mu_i,\cdots,\mu_{k} \rangle ; \mu_i = 0, 1,\ldots d-1 \}$, which is an orthonormal basis of Hilbert space with respect to the inner product of ${\cal H}^{\otimes k}$ defined by
\begin{equation}
    \langle \mu_1, \mu_2,...,\mu_{k}\vert  \mu'_1, \mu'_2,...,\mu'_{k} \rangle = \delta_{\mu_1 \mu'_1} \delta_{\mu_2 \mu'_2}.... \delta_{\mu_{k} \mu'_{k}}.
\end{equation}
According to the approach developed \cite{Vourdas1990} and mainly studied \cite{Ellinas1992,sanders1995}, we decompose the qubit raising and lowering operators as
\begin{equation} 
\mathcal{B}^{-}_{i}=\hat{E}_i ~\sqrt{\mathcal{B}^{+}_{i}\mathcal{B}^{-}_{i}}  ~~~\qquad~~~ \mathcal{B}^{+}_{i}=  \sqrt{\mathcal{B}^{+}_{i}\mathcal{B}^{-}_{i}}~ \hat{E}^\dagger_i, 
\end{equation}

The phase operators $\hat{E}_i$ are pairwise commutative and can be simultaneously diagonalized in the same basis, which facilitates their analysis. We are now ready to derive the eigenstates of the $\hat{E}_i$ operator. To do this, we solve the following eigenvalue equation
\begin{equation}
\hat{E}_{i}\vert \kappa_i \rangle= \kappa_i \vert \kappa_i \rangle. \label{ez=zz}
\end{equation}
Consequently, the complex variables $\kappa_i$ are subject to the zero power condition $\kappa_i^{d}=1$. So the number $\kappa_i$ takes discrete values on the unit circle
\begin{equation}\label{eq005}
	\kappa_i =  \Omega^{n_i}, \quad  n_i= 0,1,\ldots d-1.
\end{equation}
The common eigenbasis of all phase operators $\hat{E}_i$ is given by the product states $\vert \kappa_1 \rangle \otimes \vert \kappa_{2} \rangle \otimes \cdots \otimes \vert\kappa_{k} \rangle $. By imposing the normalization condition of the common eigenstates $\vert \kappa_i \rangle$, the $k$-qubit phase states denoted now $\vert \phi_{n_1,n_2,\cdots,n_k} \rangle$  are la-belled by the set of parameters $ ( n_1,n_2,\cdots,n_{k})$ and can be expressed as a superposition of product qudit states 
\begin{equation}\label{eq014}
	\vert \phi_{n_1,n_2,\cdots,n_k} \rangle= \frac{1}{\sqrt{d^k}}\sum_{\mu_1,\mu_2,\cdots,\mu_k} \Omega^{n_1 \mu_1}\Omega^{n_2
		\mu_2}\cdots \Omega^{n_k \mu_k}
	\vert \mu_1,\mu_2,\cdots,\mu_{k} \rangle.
\end{equation}
The corresponding $k$-qubit separable density matrices $\sigma^{\vec{n}}_{k}$  labelled by the multi-index $ \vec{n} \equiv ( n_1,n_2,\cdots,n_{k})$, write
\begin{equation}\label{mat01}
	\sigma^{\vec{n}}_{k}   = \vert \phi_{n_1,n_2,\cdots,n_k} \rangle \langle \phi_{n_1,n_2,\cdots,n_k} \vert = \frac{1}{{{d^k}}} \sum_{\mu_1,\dots,\mu_k\atop \mu^{\prime}_1,\dots,\mu^{\prime}_k} \Omega^{ \sum_{i} n_i (\mu_i - \mu'_i)}
	|\mu_1,\dots,\mu_k\rangle\langle \mu^{\prime}_1,\dots,\mu^{\prime}_k|.
\end{equation}
We note that the labeled phase density matrix (\ref{mat01}) can be expressed as
\begin{equation}\label{mat02}
	\sigma^{\vec{n}}_{k} = (\mathcal{Z}_1^{n_1}\otimes \mathcal{Z}_2^{n_2}\dots \mathcal{Z}_k^{n_k})  \sigma_{k}  (\mathcal{Z}_1^{n_1}\otimes \mathcal{Z}_2^{n_2}\dots \mathcal{Z}_k^{n_k})^{\dag}.
\end{equation}
By means of the unitary phase operators $\mathcal{Z}_i$, referred to as generalized Pauli operators acting on a single $d$-dimensional Hilbert space $ \mathcal{H}_i$ corresponding to the $i$th qudit,
\begin{eqnarray}
	\mathcal{Z}_i = \sum_{\mu_i=0}^{d-1} \Omega^{\mu_i} \vert \mu_i \rangle \! \langle \mu_i \vert  \label{eqn1}
\end{eqnarray}
The local operations $\mathcal{Z}_i$ do not affect the separability property of the state $\sigma_{k}$ given by
\begin{equation}\label{10}
	\sigma_{k} = \vert + , k \rangle \langle + , k \vert =   \left[ \frac{1}{d} \sum^{1}_{\mu_i,\mu'_i}\vert  \mu_i \rangle \langle \mu'_i \vert \right]^{\otimes k}.
\end{equation}
\section{Multipartite entangled  multi-qudit phase states} \label{sec4}
The evolved  states can be obtained from the dynamical evolution of the factor-able states (\ref{10}). This dynamical evolution write
\begin{equation}\label{vp0}
	\mathcal{U} \sigma_{k}\mathcal{U}^{\dagger} \equiv \varrho_{k}
\end{equation}
The unitary evolution operator $\mathcal{U}$ can be factorized as follows
\begin{equation}
    \mathcal{U} = \prod_{i < j} (\mathcal{C}_{ij})^{Q_{ij}}, \quad \mathcal{C}_{ij}\vert \mu_i,\mu_j \rangle = \Omega^{\mu_i \mu_j }\vert \mu_i,\mu_j \rangle
\end{equation}
with $\mathcal{C}_{ij}$ stands for the Ising type unitary operator acting on the qudits $i$ and $j$.\\

The quantities $Q_{ij}$ denote  the number of connections linking the qudits $i$ and $j$. The entangled evolved phase state $\varrho_{k}$, is written as
\begin{equation}\label{16}
	\varrho_{k} =    \frac{1}{{{d^k}}} \sum_{\mu_1,\dots,\mu_k\atop \mu^{\prime}_1,\dots,\mu^{\prime}_k}  \Omega^{(\sum_{i<j} Q_{ij}( \mu_i \mu_j- \mu'_i \mu'_j)) }
	|\mu_1,\dots,\mu_n\rangle\langle \mu^{\prime}_1,\dots,\mu^{\prime}_k|
\end{equation}
The $k$-qudit evolved  states can be generated by performing the unitary operator $\prod\limits_{i < j} {{{\left( {{\mathcal{C}_{ij}}} \right)}^{{Q_{ij}}}}}$ on the fully separable quantum  states (\ref{10}), one gets
\begin{equation}\label{20}
	\varrho_{k} = {\left(\prod\limits_{i < j} {{{\left( {{\mathcal{C}_{ij}}} \right)}^{{Q_{ij}}}}}\right)}{\sigma_k}{\left( {\prod\limits_{i < j} {{{\left( {{\mathcal{C}_{ij}}} \right)}^{{Q_{ij}}}}} } \right)^\dag }
\end{equation}
The entangled density matrices  $\varrho $  can be expressed as
\begin{equation}\label{eqbe01}
	\varrho  = \vert\psi_k \rangle\langle \psi_k \vert, \quad \vert\psi_k\rangle =  \frac{1}{{{\sqrt{d^k}}}} \sum_{\mu_1,\dots,\mu_k}
	\Omega^{\sum_{i < j} Q_{ij} \mu_i \mu_j} \vert \mu_1,\mu_2,\cdots,\mu_{k} \rangle
\end{equation}
where \( \vert \psi_k \rangle \) is the vector multi-qudit entangled state. To study the bipartite entanglement of the resulting entangled multi-qudit system, we divide the whole system into two subsets \( \big[\mathcal{A}_{2} = \mu_1, \ldots, \mu_{\lfloor \frac{k}{2} \rfloor} \} \big] \bigcup \big[\mathcal{A}_{1} = \mu_{\lfloor \frac{k}{2} \rfloor + 1}, \ldots, \mu_k \} \big] \). The matrix representing the connections between the qudits living in \( \mathcal{A}_{1} \) and the qudits of \( \mathcal{A}_{2} \) is given by
\begin{equation}
    \mathcal{C}_\mu=
\begin{pmatrix}
\vartheta_{1(\lfloor \frac{k}{2}\rfloor+1)} & \vartheta_{1(\lfloor \frac{k}{2}\rfloor+2)} & \ldots &\ldots & \vartheta_{1(k-1)} & p_{1k}\\
\vartheta_{2(\lfloor \frac{k}{2}\rfloor+1)} & \vartheta_{2(\lfloor \frac{k}{2}\rfloor+2)} & \ldots &\ldots & \vartheta_{2(k-1)} & \vartheta_{2k}\\
\vdots & \ldots& \ldots&\ldots & \ldots &\vdots\\
\vdots & \ldots& \ldots&\ldots & \ldots &\vdots\\
\vartheta_{(\lfloor \frac{k}{2}\rfloor-1)(\lfloor \frac{k}{2}\rfloor+1)} & \vartheta_{(\lfloor \frac{k}{2}\rfloor-1)(\lfloor \frac{k}{2}\rfloor+2)} & \ldots &\ldots & \vartheta_{(\lfloor \frac{k}{2}\rfloor-1)(k-1)} & \vartheta_{(\lfloor \frac{k}{2}\rfloor-1)k}\\
\vartheta_{\lfloor \frac{k}{2}\rfloor(\lfloor \frac{k}{2}\rfloor+1)} & \vartheta_{\lfloor \frac{k}{2}\rfloor(\lfloor \frac{k}{2}\rfloor+2)} & \ldots &\ldots & \vartheta_{\lfloor \frac{k}{2}\rfloor(k-1)} & \vartheta_{\lfloor \frac{k}{2}\rfloor k}\\
\end{pmatrix}
\end{equation}
The density matrix of the subsystem $\mathcal{A}_{2}$ is given by $\varrho_{\mathcal{A}_{2}}=Tr_{\mathcal{A}_{1}}[|\psi_k\rangle\langle\psi_k|]$, with $|\psi_k\rangle $ is a maximally entangled state, with respect to the bipartition of the entire system into the two parts $[\mathcal{A}_{2}=\{\mu_1,\ldots,\mu_{\lfloor \frac{k}{2}\rfloor}\}]\bigcup [\mathcal{A}_{1}=\{\mu_{\lfloor \frac{k}{2}\rfloor+1},\ldots, \mu_{k}\}]$, that is, the von Newmann entropy is maximal,
\begin{equation}
    S(\varrho_{\mathcal{A}_{2}})=  \lfloor \frac{k}{2}\rfloor ~ \log_2~ d,
\end{equation}
If, and only if, the vectors \( \mu \) \( (\vartheta_{i(\lfloor \frac{k}{2} \rfloor + 1)}, \vartheta_{i(\lfloor \frac{k}{2} \rfloor + 2)}, \ldots, \vartheta_{ik}) \) with \( (1 \leq i \leq \lfloor \frac{k}{2} \rfloor ) \) are linearly independent \cite{Helwig2012}. In the following we termed \( | \psi_{\left( {\lfloor \frac{k}{2} \rfloor, k} \right)} \rangle \) PME state with respect to the bipartition \( [\mathcal{A}_{2} = \{\mu_1, \ldots, \mu_{\lfloor \frac{k}{2} \rfloor} ] \bigcup [\mathcal{A}_{1} = \{\mu_{\lfloor \frac{k}{2} \rfloor + 1}, \ldots, \mu_{k} \}] \) and is expressed as 
\begin{equation} 
	|{\psi _{\left( {\lfloor \frac{k}{2}\rfloor,k} \right)}}\rangle =  \frac{1}{{{\sqrt{d^k}}}} \sum_{\mu_1,\dots,\mu_k}
	\Omega^{\sum_{i < j} Q_{ij} \mu_i \mu_j} \vert \mu_1,\mu_2,\cdots,\mu_{k} \rangle.  \label{a1}
\end{equation}
\section{PME states from phase states} \label{sec5}
We can now extract PME states from entangled multi-qudit phase states. The bipartite entanglement of the intricate multi-qudit system is studied in terms of dividing the whole system into two subsets \( \big[\mathcal{A}_{2} \big] \bigcup \big[\mathcal{A}_{1} \big] \) such that \( |\mathcal{A}_{2}| = \lfloor \frac{k}{2} \rfloor \leq |\mathcal{A}_{1}| = k - \lfloor \frac{k}{2} \rfloor \). If each bipartition of adjacent particles that satisfy the periodic boundary condition is completely mixed, then the entangled phase state is a PME state.
\subsection{Bipartite PME states}
For a bipartite qubit system, the maximally entangled clustered state (\ref{20}), for $K = 2$, is given by
\begin{equation}
	{\varrho_2} = \frac{1}{{{2^2}}}\sum\limits_{\scriptstyle{\mu_1},{\mu_2}\hfill\atop
		\scriptstyle \mu_1^,,\mu_2^,\hfill} {{{\left( {\Omega} \right)}^{\left( {{\mu_1}{\mu_2} - \mu_1^,\mu_2^,} \right)}}\left| {{\mu_1},{\mu_2}} \right\rangle \left\langle {\mu_1^,,\mu_2^,} \right|} \label{24}
\end{equation}
The bipartite labeled entangled phase state, $|{\phi _{{n_1},{n_2}}}\rangle  = \frac{1}{{\sqrt {{d^2}} }}\sum\limits_{{n_1},{n_2}, \cdots ,{\mu_k}} {{{(\Omega)}^{{n_1}{n_1}}}} {(\Omega)^{{n_2}{\mu_2}}}|{\mu_1},{\mu_2}\rangle$ that is local unitary equivalent to the entangled phase state
\begin{equation}
    \vert \phi_{n_1,n_2} \rangle=  (\mathcal{Z}^{n_1}_1\otimes \mathcal{Z}^{n_2}_2 )\vert \psi_p , 2 \rangle
\end{equation}
which  is locally equivalent to the maximally entangled
two-qudit state of GHZ type clearly to the two-qudits AME states $\vert \psi_p , 2 \rangle$  can given by $\left| {\mathcal{AME}\left( {2,d} \right)} \right\rangle  = \frac{1}{{\sqrt d }}\sum\limits_{{\mu_1}}^{d - 1} {\left| {{\mu_1}} \right\rangle } \left| {{\phi _{{\mu_1}}}} \right\rangle $ where the states  $\left| {{\phi _{{\mu_1}}}} \right\rangle  $ is defined by 
\begin{equation}
    \left| {{\phi _{{\mu_1}}}} \right\rangle  = \frac{1}{{\sqrt d }}\sum\limits_{{\mu_2}} {{{( \Omega)}^{{\mu_1}{\mu_2}}}\left| {{\mu_2}} \right\rangle}
\end{equation}
The bipartite PME state can be written
\begin{equation}
    \left| {\mathcal{PME}\left( {2,d} \right)} \right\rangle  \equiv \left| {{\psi _p},2} \right\rangle  = \mathcal{CP}_{12}\frac{1}{d}\sum\limits_{{\mu_1},{\mu_2}} {\left| {{\mu_1},{\mu_2}} \right\rangle }
\end{equation}
and more developed explicitly as
\begin{equation}
    \left| {\mathcal{PME}\left( {2,d} \right)} \right\rangle  = \frac{1}{{\sqrt {{d^2}} }}\sum\limits_{{\mu_1},{\mu_2}} {{{\left( {\Omega} \right)}^{{\mu_1}{\mu_2}}}} \left| {{\mu_1},{\mu_2}} \right\rangle
\end{equation}
where ${\vartheta_{12}} = 1$ number of connection linking $1$ et $2$ qudits and the controlled phase gate is giving by 
\begin{equation}
    \mathcal{CP}_{ij} = {\sum\limits_{\mu = 0}^{d - 1} {\left| \mu \right\rangle \left\langle \mu \right|} _j} \otimes \mathcal{Z}_j^\mu
\end{equation}
if we split the two qudits into two planar groups ${\mathcal{A}_{1}}$ and ${\mathcal{A}_{2}}$ where each group is a connected
partition with $\left|{{\mathcal{A}_{1}}}\right| = \left| {{\mathcal{A}_{2}}} \right|=1$ one can show that the density matrix derived from $\left| {\mathcal{PME}\left( {2,d} \right)} \right\rangle \left\langle {\mathcal{PME}\left( {2,d} \right)} \right|$ tracing out the first qudit (${\mu_1}$) is proportional to the identity

\begin{equation}
    {\varrho _{{\mathcal{A}_{2}}}} = \left| {\mathcal{PME}\left( {2,d} \right)} \right\rangle \left\langle {\mathcal{PME}\left( {2,d} \right)} \right| = \frac{1}{d}{\mathcal{I}_d}
\end{equation}
We can also check that ${\varrho _{{\mathcal{A}_{1}}}} = \frac{1}{d} \mathcal{I}_d$ by tracing out the qudit $({\mu_2})$, then ${\mathcal{A}_{1}}$ is in a maximally mixed state, i.e. a collection of adjacent particles in a state of complete mixing. Thus, the $\left| \mathcal{PME}(2, d) \right\rangle$ state is maximally mixed for the only possible bipartition of the bipartite qudit system.
\subsection{Tripartite PME states}
For three qubits systems $(K = 3)$ and similarly to the two-qudit case discussed in the previous subsection, the entangled phase state (\ref{16}) can be written for $(Q_{12} = Q_{23} = Q_{13} = 1)$
\begin{equation}
	\varrho _3^{\left\{ {123} \right\}} = \frac{1}{{{2^3}}}\sum\limits_{\scriptstyle{\mu_1},{\mu_2},{\mu_3}\hfill\atop
		\scriptstyle \mu_1^,,\mu_2^,,\mu_3^,\hfill} {{{\left( { \Omega} \right)}^{\left( {{\mu_1}{\mu_2} - \mu_1^,\mu_2^,} \right)}}{{\left( {\Omega} \right)}^{\left( {{\mu_1}{\mu_3} - \mu_1^,\mu_3^,} \right)}}{{\left( {\Omega} \right)}^{\left( {{\mu_2}{\mu_3} - \mu_2^,\mu_3^,} \right)}}\left| {{\mu_1},{\mu_2},{\mu_3}} \right\rangle \left\langle {\mu_1^,,\mu_2^,,\mu_3^,} \right|}. \label{51}
\end{equation}
The tripartite labeled entangled phase state $|{\phi _{{n_1},{n_2},{n_3}}}{\rangle _\vartheta} = (\mathcal{Z}_1^{{n_1}} \otimes \mathcal{Z}_2^{{n_2}} \otimes \mathcal{Z}_3^{{n_3}})|{\psi _\vartheta},3\rangle$, and the tripartite PME state for qudits system writing as 
\begin{equation}
    \left| {\mathcal{PME}\left( {3,d} \right)} \right\rangle  \equiv \left| {{\psi _\vartheta},3} \right\rangle  = {\left( {\mathcal{CP}_{12}} \right)^{{\vartheta_{12}}}}{\left( {\mathcal{CP}_{23}} \right)^{{\vartheta_{23}}}}{\left( {\mathcal{CP}{13}} \right)^{{\vartheta_{13}}}}\frac{1}{d}\sum\limits_{{\mu_1},{\mu_2},{\mu_3}} {\left| {{\mu_1},{\mu_2},{\mu_3}} \right\rangle },
\end{equation}
which can be rewritten in a clear and detailed manner in the case where ${\vartheta_{12},\vartheta_{13},\vartheta_{23}} = 1$ as
\begin{equation}
    \left| {\mathcal{PME}\left( {3,d} \right)} \right\rangle  = \frac{1}{{\sqrt {{d^3}} }}\sum\limits_{{\mu_1},{\mu_2},{\mu_3}} {{{\left( {\Omega} \right)}^{{\mu_1}{\mu_2}}}{{\left( {\Omega} \right)}^{{\mu_1}{\mu_3}}}{{\left( {\Omega} \right)}^{{\mu_2}{\mu_3}}}} \left| {{\mu_1},{\mu_2},{\mu_3}} \right\rangle
\end{equation}
It is clear that this PME state is local unitary equivalent to GHZ-type states for three-dimensional systems. It can easily be shown that the density matrix derived from the state \( \left| \mathcal{PME}(3,d) \right\rangle \left\langle \mathcal{PME}(3,d) \right| \) by plotting adjacent sites in the set \( \mathcal{A}_{2} \) (qudits 1 and 2) is proportional to the identity, i.e.
\begin{equation}
    {\varrho _{{\mathcal{A}_{1}}}} = T{r_{{\mathcal{A}_{2}}}}\left( {\left| {\mathcal{PME}\left( {3,d} \right)} \right\rangle \langle \mathcal{PME}\left( {3,d} \right)|} \right) = \frac{1}{d}{\mathcal{I}_d}
\end{equation}
\subsection{Quadripartite PME states}
For four qubits system ($K=4$) the star-shape four-qubit entangled phase state $\varrho_4^{\{12,13,14 \}}$  can be written for ($Q_{12} = Q_{13} = Q_{14} =1$) as 
\begin{equation}
	\varrho _4^{\left\{ {12,13,14} \right\}} = \frac{1}{{{2^4}}}\sum\limits_{\scriptstyle{\mu_1},{\mu_2},{\mu_3},{\mu_4}\hfill\atop
		\scriptstyle \mu_1^,,\mu_2^,,\mu_3^,,\mu_4^,\hfill} {{\left( {\Omega} \right)}^{\left( {{\mu_1}{\mu_2} - \mu_1^,\mu_2^,} \right)}}{{\left( {\Omega} \right)}^{\left( {{\mu_1}{\mu_3} - \mu_1^,\mu_3^,} \right)}}{{\left( { \Omega} \right)}^{\left( {{\mu_1}{\mu_4} - \mu_1^,\mu_4^,} \right)}}
	\left| {{\mu_1},{\mu_2},{\mu_3},{\mu_4}} \right\rangle \left\langle {\mu_1^,,\mu_2^,,\mu_3^,,\mu_4^,} \right|
\end{equation}
For the second case given by the line-shape four-qubit entangled phase state $\varrho_4^{\{12,23,34 \}}$ given by
\begin{equation}
     \varrho _4^{\left\{ {12,23,34} \right\}} = \frac{1}{{{2^4}}}\sum\limits_{\scriptstyle{\mu_1},{\mu_2},{\mu_3},{\mu_4}\hfill\atop\scriptstyle \mu_1^,,\mu_2^,,\mu_3^,,\mu_4^,\hfill} {{\left( { \Omega} \right)}^{\left( {{\mu_1}{\mu_2} - \mu_1^,\mu_2^,} \right)}}{{\left( {\Omega} \right)}^{\left( {{\mu_2}{\mu_3} - \mu_2^,\mu_3^,} \right)}}{{\left( { \Omega} \right)}^{\left( {{\mu_3}{\mu_4} - \mu_3^,\mu_4^,} \right)}}
     \left| {{\mu_1},{\mu_2},{\mu_3},{\mu_4}} \right\rangle \left\langle {\mu_1^,,\mu_2^,,\mu_3^,,\mu_4^,} \right|
\end{equation}
The four-qudit entangled labeled phase states are local unitary equivalent to the entangled phase states shared between four-qudits and task the form
\begin{equation}
    |{\phi _{{n_1},{n_2},{n_3},{n_4}}}{\rangle _\vartheta} = (\mathcal{Z}_1^{{n_1}} \otimes \mathcal{Z}_2^{{n_2}} \otimes \mathcal{Z}_3^{{n_3}}\otimes \mathcal{Z}_2^{{n_4}})|{\psi _\vartheta},4\rangle
\end{equation}
The entangled phase states of (23) can be expressed in terms of the controlled phase gates $\mathcal{CP}$ as
\begin{equation}
   \left| {\mathcal{PME}\left( {4,d} \right)} \right\rangle  \equiv \left| {{\psi _\vartheta},4} \right\rangle  = \left( {\prod\limits_{j > i}^4 {{{\left( {\mathcal{CP}_{ij}} \right)}^{{\vartheta_{ij}}}}} } \right)\frac{1}{{\sqrt {{d^4}} }}\sum\limits_{{\mu_1},{\mu_2},{\mu_3},{\mu_4}} {\left| {{\mu_1},{\mu_2},{\mu_3},{\mu_4}} \right\rangle }
\end{equation}
From the definition in \ref{def2}, the quadripartite PME state is
\begin{equation}
		\left| {\mathcal{PME}\left( {4,d} \right)} \right\rangle  = \prod\limits_{l = 1}^2 {{{\left| {{\mathcal{P}^ + }} \right\rangle }_{l,l + 2}}}= {\left| {{\mathcal{P}^ + }} \right\rangle _{1,3}}.{\left| {{\mathcal{P}^ + }} \right\rangle _{2,4}}
\end{equation}
Consider an arbitrary bipartition into ${\mathcal{A}_{2}} = \left\{ {{\mu_3},{\mu_4}} \right\}$ and ${\mathcal{A}_{1}} = \left\{ {{\mu_1},{\mu_2}} \right\}$ with $\left| {{\mathcal{A}_{2}}} \right| = \left| {{\mathcal{A}_{1}}} \right|=2$ which can be rewritten as
\begin{equation}
    \left| {\mathcal{PME}\left( {4,d} \right)} \right\rangle  = \mathcal{CP}_{13}\left( {\frac{1}{{\sqrt d }}\sum\limits_{i = 0}^{d - 1} {\left| {i,i} \right\rangle }} \right)\mathcal{CP}_{24}\left( {\frac{1}{{\sqrt d }}\sum\limits_{i = 0}^{d - 1} {\left| {j,j} \right\rangle }} \right)
\end{equation}
we note that the density matrix of four qudits state reads $${\varrho _{{\mathcal{A}_{1}}{\mathcal{A}_{2}}}} = \left| {\mathcal{PME}\left( {4,d} \right)} \right\rangle \left\langle {\mathcal{PME}\left( {4,d} \right)} \right|$$
to derive the density matrix associated to the subsystem ${\mathcal{A}_{2}} = \left\{ {{\mu_3},{\mu_4}} \right\}$  , we
have to consider the reduced state ${\varrho _{{\mathcal{A}_{2}}}} = T{r_{{\mathcal{A}_{1}}}}\left( {\left| {\mathcal{PME}\left( {4,d} \right)} \right\rangle \langle \mathcal{PME}\left( {4,d} \right)|} \right)$ wite we found that is totally mixed
\begin{equation}
    {\varrho _{{\mathcal{A}_{2}}}} = T{r_{{\mathcal{A}_{1}}}}\left( {\left| {\mathcal{PME}\left( {4,d} \right)} \right\rangle \langle \mathcal{PME}\left( {4,d} \right)|} \right) = \frac{1}{{{d^{_{\left| {{\mathcal{A}_{2}}} \right|}}}}}{I_d}^{^{_{\left| {{\mathcal{A}_{2}}} \right|}}}
\end{equation}
\subsection{Eight-partite PME states}
In this case the five qubits system $(K=8)$, the star-shape eight-qubit entangled phase state $\varrho_8^{\{12,13,14,15,16,17,18 \}}$  can be written for ($Q_{12} = Q_{13} = Q_{14} =Q_{15}=Q_{16}=Q_{17}=Q_{18}=1$) as
\begin{align}
    \varrho _5^{\left\{ {12,13,14,15,16,17,18} \right\}} &= \frac{1}{{{2^8}}}\sum\limits_{\scriptstyle{\mu_1},{\mu_2},{\mu_3},{\mu_4},{\mu_5},{\mu_6},{\mu_7},{\mu_8}\hfill\atop
	\scriptstyle \mu_1^,,\mu_2^,,\mu_3^,,\mu_4^,,\mu_5^,,\mu_6^,,\mu_7^,,\mu_8^,\hfill} {{{\left( { \Omega} \right)}^{\left( {{\mu_1}{\mu_2} - \mu_1^,\mu_2^,} \right)}}{{\left( { \Omega} \right)}^{\left( {{\mu_1}{\mu_3} - \mu_1^,\mu_3^,} \right)}}{{\left( {\Omega} \right)}^{\left( {{\mu_1}{\mu_4} - \mu_1^,\mu_4^,} \right)}}
	\left( { \Omega} \right)}^{\left( {{\mu_1}{\mu_5} - \mu_1^,\mu_5^,} \right)} {\left( { \Omega} \right)}^{\left( {{\mu_1}{\mu_6} - \mu_1^,\mu_6^,} \right)}\\
  &{\left( { \Omega} \right)}^{\left( {{\mu_1}{\mu_7} - \mu_1^,\mu_7^,} \right)}{\left( { \Omega} \right)}^{\left( {{\mu_1}{\mu_8} - \mu_1^,\mu_8^,} \right)}
\left|{{\mu_1},{\mu_2},{\mu_3},{\mu_4},{\mu_5},{\mu_6},{\mu_7},{\mu_8}} \right\rangle \left\langle {\mu_1^,,\mu_2^,,\mu_3^,,\mu_4^,,\mu_5^,,\mu_6^,,\mu_7^,,\mu_8^,} \right|
\end{align}
The eight-partite PME is giving by
\begin{equation}
    \left| {\mathcal{PME}\left( {8,d} \right)} \right\rangle  = {\kern 1pt} \frac{1}{{\sqrt {{d^8}} }}\sum\limits_{{\mu_1},{\mu_2},{\mu_3},{\mu_4},{\mu_5},{\mu_6},{\mu_7},{\mu_8}} {{{\left( { \Omega} \right)}^{{\mu_1}{\mu_2}}}{{\left( { \Omega} \right)}^{{\mu_1}{\mu_3}}}{{\left( { \Omega} \right)}^{{\mu_1}{\mu_4}}}{{\left( { \Omega} \right)}^{{\mu_1}{\mu_5}}}} {\left( { \Omega} \right)^{{\mu_1}{\mu_6}}}{\left( { \Omega} \right)^{{\mu_1}{\mu_7}}}{\left( {\Omega} \right)^{{\mu_1}{\mu_8}}}	\left| {{\mu_1},{\mu_2},{\mu_3},{\mu_4},{\mu_5},{\mu_6},{\mu_7},{\mu_8}} \right\rangle
\end{equation}
By use the definition \ref{def2} The eight-partite PME  $(K = 2k = 8)$ is reads as
\begin{equation}
    \left| {\mathcal{PME}\left( {8,d} \right)} \right\rangle  = \prod\limits_{l = 1}^4 {{{\left| {{\mathcal{P}^ + }} \right\rangle }_{l,l + 4}}} = {\left| {{\mathcal{P}^ + }} \right\rangle _{1,5}}.{\left| {{\mathcal{P}^ + }} \right\rangle _{2,6}}.{\left| {{\mathcal{P}^ + }} \right\rangle _{3,7}}.{\left| {{\mathcal{P}^ + }} \right\rangle _{4,8}}
\end{equation}
has been developed in terms of a controlled phase gate as a
\begin{equation}
    \left| {\mathcal{PME}\left( {8,d} \right)} \right\rangle  = \mathcal{CP}_{15}.\mathcal{CP}_{26}.\mathcal{CP}_{37}.\mathcal{CP}_{48}\frac{1}{{\sqrt {{d^8}} }}\sum\limits_{i,j,k,l}{\left| {i,j,k,l,i,j,k,l} \right\rangle }
\end{equation}
\begin{center}
\begin{figure}[h]
    \centering
    \includegraphics[scale=0.9]{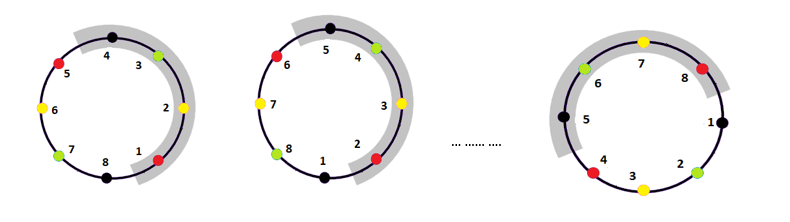}
    \caption{Illustration of an eight-particle Planar Maximally Entangled (PME) state}
    
\end{figure}
\end{center}
where $\left\lfloor {\frac{K}{2}} \right\rfloor  = \frac{8}{2} = 4$ colored area (highlighted in gray) shows a subset that contain one half of four adjacent connected particles as Bell pair is in completely mixed state. as the state in  the equation (52) represent  we show Bell pairs represented by matching colors for example particles 1 and 5 (in red), particles 2 and 6 (yellow), particles 3 and 7 (green), and particles 4 and 8 (black). This pairing  corresponds to the Controlled-Phase (CP) which establish entanglement between these pairs 
   
\subsection{Multipartite PME states}
The construction developed above can be generalised to the case of $K$-qudits. In this case, the intricate phase states expressed in terms of CP actions of the controlled phase gates are as follows
\begin{equation}
    \left| {{\psi _\vartheta},K} \right\rangle  = \left( {\prod\limits_{j > i}^K {{{\left( {{\mathcal{CP}_{ij}}} \right)}^{{\vartheta_{ij}}}}} } \right)\frac{1}{{\sqrt {{d^K}} }}\sum\limits_{{\mu_1},{\mu_2},{\mu_3},....,{\mu_K}} {\left| {{\mu_1},{\mu_2},{\mu_3}...{\mu_K}} \right\rangle }
\end{equation}
Where the entangled labeled phase states of $K$-qudit are equivalent under unitary transformation to the entangled phase state shared between $K$-qudits,
\begin{equation}
    |{\phi _{{n_1},{n_2},....,,{n_K}}}{\rangle _\vartheta} = (\mathcal{Z}_1^{{n_1}} \otimes \mathcal{Z}_2^{{n_2}} ....\otimes \mathcal{Z}_K^{{n_K}})|{\psi_\vartheta},K\rangle
\end{equation}
PME states in arbitrary dimension for $K=2k$ parties gives as
\begin{equation}
    \left| {\mathcal{PME}\left( {K,d} \right)} \right\rangle  = \prod\limits_{l = 1}^k {{{\left| {{\mathcal{P}^+}} \right\rangle }_{l,l + k}} = \prod\limits_{l = 1}^{\frac{K}{2}} {{{\left| {{\mathcal{P}^+}}\right\rangle}_{l,l+\frac{K}{2}}}}},
\end{equation}
\begin{equation}
    \left|{\mathcal{PME}\left( {K,d} \right)} \right\rangle  = {\left| {{\mathcal{P}^+}} \right\rangle _{1,1 + \frac{K}{2}}}.{\left| {{\mathcal{P}^ + }} \right\rangle _{2,2 + \frac{K}{2}}} ....{\left| {{\mathcal{P}^ + }} \right\rangle _{\frac{K}{2},K}},
\end{equation}
\begin{equation}
    \left| {\mathcal{PME}\left( {K,d} \right)} \right\rangle  = \mathcal{CP}_{1,1 + \frac{K}{2}} ............\mathcal{CP}_{\frac{K}{2},K}\frac{1}{{{{\left( {\sqrt d } \right)}^{\frac{K}{2}}}}}\sum\limits_{{\mu_1},....,{\mu_2}} {\left| {{\mu_1},{\mu_2},.........,{\mu_{\frac{K}{2}}}} \right\rangle}.
\end{equation}
The density matrix $\varrho_{\mathcal{A}_{1}\mathcal{A}_{2}}$ associated with the $K$-qudit PME state is as follows
\begin{equation}
    {\varrho _{{\mathcal{A}_{1}}{\mathcal{A}_{2}}}} = \left| {\mathcal{PME}\left( {K,d} \right)} \right\rangle \left\langle {\mathcal{PME}\left( {K,d} \right)} \right|
\end{equation}
The reduced density matrix of the subset $\mathcal{A}_{2}$ which contains half of the adjacent qubits is totally mixed,
\begin{equation}
    {\varrho _{{\mathcal{A}_{2}}}} = T{r_{{\mathcal{A}_{1}}}}\left( {\left| {\mathcal{PME}\left( {K,d} \right)} \right\rangle \langle \mathcal{PME}\left( {K,d} \right)|} \right) = \frac{1}{{{d^{_{\left| {{\mathcal{A}_{2}}} \right|}}}}}{\mathcal{I}_d}^{^{_{\left| {{\mathcal{A}_{2}}} \right|}}}.
\end{equation}
\section{Quantum Secret sharing with PME multi-qudit states} \label{sec6}
We demonstrate in this section how a PME state can
be used to construct a quantum secret sharing scheme \cite{Helwig2013,Markham2010,Keet2010}. Thus, in a QSS protocol, a dealer encodes a secret into a quantum state that is shared among \( k \) players such that only special subsets of players are able to recover the secret. PME states can be used for QSS between \( 2k-1 \) players so that any subset of adjacent players larger than \( k \) can recover the state. The players must be adjacent to each other, i.e. they must form a connected subset of the set of players.

In this context, the topological structure of the players plays a crucial role. Adjacent players must be able to interact with each other to ensure that the information contained in the shared quantum state can be recovered correctly. This connectivity condition is essential for the security of secret sharing, as it ensures that non-adjacent players, not having access to information from other intermediate players, cannot deduce the secret. Thus, the use of PME states in a QSS protocol not only ensures the quantum security of information, but also guarantees flexibility in the organization of subsets of players, according to the needs of the application. In the following, we briefly describe threshold quantum secret sharing schemes associated with a maximally intricate $\left( {\lfloor \frac{k}{2}\rfloor,k} \right)$ planar state shared between \( K \) qudits ${\rm{\{ }}{{\rm{\mu}}_1}{\rm{, }}...,{\rm{ }}{{\rm{\mu}}_K}{\rm{\} }} $. The secret to be shared is a quantum state $\left| \lambda  \right\rangle$ in a Hilbert space of $d$ dimensions, initially possessed by the dealer, who distributes it to the other parties. This secret is defined by an arbitrary normalized quantum state expressed by
\begin{equation}
   \left| \lambda  \right\rangle  = \sum\limits_{i = 1}^{d - 1} {{\lambda _i}\left| i \right\rangle }
\end{equation}
we choose the dealer $D$ assigned to one of the qudits of the multi-qudit system and to be the first player $\left\{ {{\mu_1}} \right\}$ of the set $\mathcal{A}_{2}$, we consider the maximally intricate planar state $\left( {\lfloor \frac{k}{2}\rfloor, k} \right)$ suitable $|{\psi _{\left( {\lfloor \frac{k}{2}\rfloor, k} \right)}}\rangle $ with respect to the bipartition $[\mathcal{A}_{2}=\mu_1,\ldots, \mu_{\lfloor \frac{k}{2}\rfloor}\}]\bigcup [\mathcal{A}_{1}=\mu_{\lfloor \frac{k}{2}\rfloor+1},\ldots, \mu_{k}\}]$ of equal size $\lfloor \frac{k}{2}\rfloor$. In the addition, the no-cloning theorem guarantees that any set of less than $\lfloor \frac{k}{2}\rfloor$ players cannot gain any information about the state. The qudits living in ${\mathcal{A}_{1}}$ and the qudits of ${\mathcal{A}_{2}}$ are connected by the interaction parameters $_{Qij}$ and they are chosen to be the legal players in this QSS scheme. The dealer prepares the state $\left( {\lfloor \frac{k}{2}\rfloor,k} \right)$ PME state by combining it with a quantum secret $\left| \lambda \right\rangle $ form a new $ \left( {k + 1} \right)$-qudit entangled state can be written as 
\begin{equation}
    \left| \lambda  \right\rangle  \otimes |{\psi _{\left( {\lfloor \frac{k}{2}\rfloor,k} \right)}}\rangle  = \left| {\xi (k + 1)} \right\rangle
\end{equation}
Here, each of the players performs a Bell measurement on their respective qudit of the PME state and a qudit of the secret with respect to the dealer, who distributes each share $\left| {{\mu_i}} \right\rangle $ to each other player, i.e. the dealer then measures her two qudits in the generalized Bell basis,
\begin{equation}
    \left| {{{\rm \mathcal{M}}_{nk}}} \right\rangle  = \frac{1}{{\sqrt d }}\sum\limits_{j = 0}^{d - 1} {{\Omega ^{jk}}\left| j \right\rangle \left| {j + n} \right\rangle}.  
\end{equation}
The resultant state for all parties is
\begin{equation}
	\left| {{{\rm \mathcal{M}}_{nk}}} \right\rangle \left\langle {{{\rm \mathcal{M}}_{nk}}} \right|\left| \lambda  \right\rangle  \otimes |{\psi _{(\frac{k}{2},k)}}\rangle  = \left| {{{\rm \mathcal{M}}_{nk}}} \right\rangle \left\langle {{{\rm \mathcal{M}}_{nk}}} \right|\left. {\xi (k+1)} \right\rangle\left| {{{\rm M}_{nk}}} \right\rangle \sum\limits_{j=0}^{d-1} {{\alpha_j} {\Omega^{-jk}}\left| {{\varphi_j}} \right\rangle}. 
\end{equation} 
In case that the dealer informs the players of their measurement result $(n, k)$ a subset of $\left( {k - \left\lfloor {\frac{k}{2}}\right\rfloor}\right)$ players can apply the following correction operator 
\begin{equation}
	O_{nk} = S_a^{ - k\vartheta_{1a}^{ - 1}}{\rm Z}_{\left\lfloor {\frac{K}{2}} \right\rfloor  + 1}^{ - n{\vartheta_{1\left( {\left\lfloor {\frac{K}{2}} \right\rfloor  + 1} \right)}}}\,...\,\,{\rm \mathcal{Z}}_K^{ - n{\vartheta_{_{1}}K}}; {\kern 1pt} {\kern 1pt} {\kern 1pt} {\kern 1pt} {\kern 1pt} {\kern 1pt} {\kern 1pt} {\kern 1pt} {\kern 1pt} {\kern 1pt} {\kern 1pt} {\kern 1pt} {\kern 1pt} {\kern 1pt} {\kern 1pt} {\kern 1pt} {\kern 1pt} {\kern 1pt} a \in \left\{ {\left\lfloor {\frac{K}{2}} \right\rfloor  + 1,{\kern 1pt} {\kern 1pt} {\kern 1pt} {\kern 1pt} ...{\kern 1pt} {\kern 1pt} {\kern 1pt} {\kern 1pt} ,K} \right\}.
\end{equation}
By applying the correction operator the players can recover the secret shared by the dealer then we obtain 
\begin{equation}\label{00}
    S_a^{ - k\vartheta_{1a}^{ - 1}}{\rm Z}_{\left\lfloor {\frac{K}{2}} \right\rfloor  + 1}^{ - n{\vartheta_{1\left( {\left\lfloor {\frac{K}{2}} \right\rfloor+1} \right)}}}{\kern 1pt} {\kern 1pt} {\kern 1pt} \,...{\kern 1pt} \,\,{\rm \mathcal{Z}}_K^{ - n{\vartheta_{_{1}}K}}\left( {\sum\limits_j {{\alpha _j}{\Omega ^{-jk}}\left|{{\varphi _j}} \right\rangle}} \right){\kern 1pt} = \sum\limits_j {{\alpha _j}\left| {\widetilde {{\varphi _j}}} \right\rangle}.
\end{equation}
Note that ${S_a}$ is a stabilizer generators defined as 
\begin{equation}
    {S_i} = {\mathcal{Y}_i}\prod\limits_{i \ne j, j = 1}^K {{\rm \mathcal{Z}}_j^{{Q_{ij}}}},
\end{equation}
where $\mathcal{Y}$ and ${\rm \mathcal{Z}}$ are the the generalized Pauli operators using in the qudit case, and they satisfy ${\mathcal{Y}^d} = {{\rm \mathcal{Z}}^d} = \mathcal{I}$. The generalized Pauli operators acting on a single d-dimensional Hilbert space as

\begin{equation}
    \mathcal{Y}\left| \mu \right\rangle  = \left| {\mu + 1} \right\rangle{\rm \mathcal{Z}}\left| \mu \right\rangle  = {\Omega ^\mu}\left| \mu \right\rangle,
\end{equation}
where  $\Omega  = {e^{2\pi i/d}}$. Finally based on the Eq. (\ref{00}) it is noted that this scheme requires the
collaboration of at least $\left( {K - \left\lfloor {\frac{K}{2}} \right\rfloor } \right) $ players specified by the dealer to access the secret.
\begin{table}[h]
    \centering
    \renewcommand{\arraystretch}{1.3}
    \setlength{\tabcolsep}{4pt}
   
    \begin{tabular}{|p{2.2cm}|p{5cm}|p{5cm}|p{5cm}|}
        \hline
        \textbf{Feature} & \textbf{GHZ-Based QSS \cite{hillery1999, ekert1991}} & \textbf{AME-Based QSS \cite{Helwig2013, Markham2010}} & \textbf{PME-Based QSS \cite{karimipour2020, slaoui2023}} \\
        \hline
        \textbf{Scalability} & Fixed number of participants & Limited in low dimensions & Flexible, adaptable for any $K$ \\
        \hline
        \textbf{Security} & Strong but vulnerable to specific attacks & Strong, any subset of $\lfloor K/2 \rfloor$ qudits is maximally mixed & Robust, connected subsets of $\lfloor K/2 \rfloor$ particles are maximally mixed \\
        \hline
        \textbf{Efficiency} & Requires multi-qubit entanglement & High entanglement cost, challenging for large $K$ & Efficient, requires only local CP operations \\
        \hline
        \textbf{Resource Requirements} & Moderate & High due to strong entanglement constraints & Low, simpler entanglement structure \\
        \hline
        \textbf{Practical Implementation} & Requires precise entanglement distribution & Difficult due to AME state constraints & Easier, as PME states are more accessible than AME states \\
        \hline
    \end{tabular}
    \caption{Comparison of PME-Based QSS with Existing QSS Protocols.}
    \label{tab:qss_comparison}
\end{table}

The table (\ref{tab:qss_comparison}) compares three different approaches to quantum secret sharing (QSS): GHZ-based, AME-based, and PME-based protocols. Each method has its own strengths and weaknesses in terms of scalability, security, efficiency, resource requirements, and ease of implementation in real-world scenarios. Regarding scalability, GHZ-based protocols impose a fixed number of participants, while AME-based protocols are limited to low dimensions. In contrast, PME-based protocols offer greater flexibility, allowing for any number $K$ of participants.\par 
In terms of security, GHZ-based protocols provide strong protection but are not immune to certain types of attacks. AME-based protocols enhance security by ensuring that no small group of participants can access the full secret, making them more robust, PME-based protocols further improve security by ensuring that any connected subset of $\lfloor K/2 \rfloor$ particles is maximally mixed, offering stronger resilience against potential threats. When it comes to efficiency, GHZ-based protocols require multi-qubit entanglement which can be complex, while AME-based protocols also have a high entanglement cost making them tricky to use for the implementation of large $K$ particles but on the other side PME-based protocols are more efficient because they require only local controlled-phase (CP) operations which are simpler to manage.\par

In terms of resource requirements, GHZ-based protocols need a moderate amount of resources, while AME-based protocols are much more demanding because of their strict entanglement needs. PME-based protocols, however, are more resource-friendly due to their simpler entanglement structure, making them easier to work with. Finally, in terms of practical implementation, GHZ states require very precise entanglement distribution, which can be challenging. AME states are even harder to implement because of their intrinsic constraints, PME states, however, are more experimentally feasible, as they are easier to generate than AME states. Consequently, PME-based protocols strike a good balance between efficiency, resource use, and practicality, making them a strong choice for many quantum secret sharing applications.

\section{Concluding remarks} \label{clc}
To conclude this article, we summarize the main results obtained in the course of our research. We defined the phase operator using a mathematical description of the qudit system in the framework of the Weyl-Heisenberg algebra. We got the comfortable phase state from this phase operator by solving the corresponding eigenvalue equation. This gave us a solid foundation for studying the quantum properties in the future. By acting the unitary operator on the qubits of the separable phase state, we also generated an intricate evolved phase state from the dynamical evolution process. This dynamic evolution process is crucial for understanding how entangled states behave under different conditions, and it opens the way to potential applications in advanced quantum technologies. In addition, we have constructed PME states in any dimension and for any number of particles. We achieved this by studying the entanglement between the bipartitions of the entangled phase states shared between two, three, four, and several qubits. This detailed analysis of entanglement has enabled us to gain a better understanding of the structure of PME states and their potential application in various quantum communication and quantum algorithmic protocols.\par

Furthermore, we have shown that entangled labeled phase states can be expressed in terms of separable phase states. Moreover, these states are locally unitary equivalent to planar entangled maximal states via a controlled phase gate operation, which preserves the amount of entanglement. This type of state is actually broader and more generalized than maximal entangled states, implying that AME states are a subclass of PME states (phase mixed entangled states). In other words, any AME state can be considered a PME state, but the reverse is not necessarily true. Because of the remarkable properties of AME states, it has been shown that they can be applied in many fields, such as quantum computation and quantum cryptography. Therefore, PME states, which share many characteristics with AME states, should also lead to relevant results in these applications. Thus, in the context of AME applications, the expression ‘any subset’ should generally be replaced by ‘connected subsets’ or subsets of adjacent particles to better reflect the properties of entanglement that are involved.\par

Finally, PME states are of significant physical importance, as they can be widely used in various quantum information tasks. This importance is closely linked to their performance in encoding quantum information and protecting against incoherence effects. In addition, these types of states are not only suitable but also widely used in applications. Their ability to maintain levels of entanglement while being robust to perturbations makes them ideal candidates for future technological developments in the field of quantum information.\par

While the theoretical foundation of PME states is robust, their practical realization faces significant hurdles that must be overcome to leverage their potential in quantum technologies. Scalability is a primary concern, as the complexity of constructing PME states grows rapidly with system size, demanding precise control over an increasing number of particles and posing substantial computational challenges. Furthermore, current quantum hardware may lack the necessary capabilities.  Limitations in gate fidelity, error rates, and qubit connectivity can hinder the creation of these complex states. Like all entangled states, PME states are highly susceptible to decoherence, making them vulnerable to environmental noise.  Experimental validation is therefore crucial to assess their performance in realistic, noisy environments. Finally, verifying the entanglement properties of PME states requires sophisticated measurement techniques capable of confirming the expected maximal mixedness in adjacent subsystems. Overcoming these challenges is essential for bridging the gap between theoretical promise and practical application. Future research should focus on developing feasible experimental protocols for constructing PME states on current quantum platforms, outlining a roadmap for overcoming the identified practical limitations.\\

{\bf Declaration of competing interest:}\par
The authors declare that they have no known competing financial interests or personal relationships that could have appeared to influence the work reported in this paper.\\

{\bf Data availability:}\par
No data was used for the research described in the article.

\end{document}